\documentclass[10pt,superscriptaddress,showpacs,preprintnumbers]{revtex4}
\usepackage{epsf} 
\usepackage{amsmath,color}
\usepackage{amssymb}
\usepackage{epsfig}
\usepackage{latexsym}
\usepackage{amsfonts}
\usepackage{varioref}
\usepackage{ifthen}
\usepackage{float}
\usepackage{graphicx}
\usepackage{amssymb,amsmath,amsthm}
\providecommand{\U}[1]{\protect\rule{.1in}{.1in}}
\begin{document}
\parindent 0mm 
\setlength{\parskip}{\baselineskip} 
\pagenumbering{arabic} 
\setcounter{page}{1}
\mbox{ }
%
\title{Hadronic contribution to the running QED coupling at the Z-boson mass scale}
%
\author{C. A. Dominguez}
\affiliation{Centre for Theoretical \& Mathematical Physics and Department of Physics, University of
	Cape Town, Rondebosch 7700, South Africa}
\author{L. A. Hernandez}
\affiliation{Departamento de F\'isica, Universidad Aut\'onoma Metropolitana-Iztapalapa, Av. San Rafael Atlixco 186, C.P, CdMx 09340, Mexico.}
\affiliation{Centre for Theoretical \& Mathematical Physics and Department of Physics, University of
	Cape Town, Rondebosch 7700, South Africa}
\affiliation{Facultad de Ciencias de la Educaci\'on, Universidad Aut\'onoma de Tlaxcala, Tlaxcala, 90000, Mexico.}

\date{\today}
\begin{abstract}
\noindent
An update is described of a model independent method to determine the hadronic contribution to the QED running coupling at the Z-boson mass scale, $\Delta\alpha_{\text{HAD}}(M_{Z}^{2})$. The major source of uncertainty is from the contribution of the light quark vector current correlator  at zero momentum. This uncertainty is substantially reduced using recently  improved lattice QCD results for this correlator. The result is  $\Delta\alpha_{\text{HAD}}(M_{Z}^{2})=274.13 (0.73)\, \times 10^{-4}$.
\end{abstract}

\maketitle
\section{Introduction}

The electromagnetic running coupling at the scale of the Z-boson mass, $\alpha(M_{Z}^{2})$, is currently not known precisely. The main reason being the uncertainty from the hadronic sector, not fully determined in perturbative QCD (PQCD). This running coupling can be written as
\begin{equation}
\alpha(s)=\frac{\alpha(0)}{1-\Delta\alpha_\text{L}(s)-\Delta\alpha_\text{HAD}(s)}\;,
\end{equation}
where $\Delta\alpha_\text{L}$ is the leptonic contribution, known precisely from perturbation theory, and $\Delta\alpha_\text{HAD}(s)$  is the hadronic counterpart. The interesting quantity is the QED coupling at the scale of the Z-boson mass, $M_Z$.  Denoting $\alpha\equiv \alpha(0)$ in the sequel, $\Delta\alpha_{\text{HAD}}(M_{Z}^{2})$ can be written as
\begin{equation}\label{eq:alpha1}
\Delta\alpha_{\text{HAD}}(M_{Z}^{2})=4 \,\pi \, \alpha\left\{\Pi(0) - Re \,[\Pi(M_{Z}^{2})]\right\}\;,
\end{equation}
where $\Pi(s)$ is the electromagnetic current correlator
\begin{eqnarray}
\Pi_{\mu\nu} (q^2) &=& i \int d^4x\,  e^{iqx} \langle 0|\, T \left(j^{\text{\,EM}}_{\mu}(x), j^{\text{\,EM}}_{\nu}(0) \right)|0\rangle \nonumber\\ 
&=& (q_\mu q_\nu - q^2 g_{\mu\nu}) \Pi(q^2) \;,
\end{eqnarray}
with  $j^{\text{EM}}_\mu(x)=\sum_f Q_f \bar{f}(x) \gamma_\mu f(x)$, and the sum is over all quark flavors $f=\{u,d,s,c,b,t\}$, with charges $Q_f$. Invoking analyticity and unitarity for $\Pi(s)$, and using the optical theorem, i.e. $R(s)=12\pi\, \text{Im}\,\Pi(s)$, where $R(s)$ is the normalized $e^+e^-$ cross-section, one can write Eq.\eqref{eq:alpha1} as a dispersion integral \cite{cabibbo1961}
\begin{equation}\label{EQ:dispersion}
\Delta\alpha_{\text{HAD}}(M_{Z}^{2})=\frac{\alpha \, M_{Z}^{2}}{3\, \pi}P\int^{\infty}_{4m_{\pi}^{2}}\frac{R(s)}{s(M_{Z}^{2}-s)}ds \;,
\end{equation}
where $P$ denotes the principal part of the integral. This expression only requires knowledge of $R(s)$, which is accessible experimentally in the resonance region, followed by perturbative QCD in the continuum.
However, given the current uncertainties in the $R(s)$ data, of diverse magnitudes depending on the energy region, several approaches  have been proposed to circumvent this issue. \\ 
The standard approach to determining $\Delta\alpha_{\text{HAD}}(M_{Z}^{2})$ is to evaluate Eq.\eqref{EQ:dispersion} making use of $e^+e^-$ annihilation data for $R(s)$ in the  resonance regions, and either use the PQCD prediction for $R(s)$ above these regions (see e.g. \cite{davier2011}), or make use of all the available $e^+e^-$ data and fill in the gaps using the PQCD prediction (see e.g. \cite{hagiwara2011, Actis}). An alternative approach was proposed in \cite{SB}, based entirely on perturbative QCD in the heavy-quark (charm and bottom) region, and Lattice QCD (LQCD) determinations of the light-quark vector current correlator at zero-momentum. The latter was only known at the time with a large uncertainty. Recent LQCD determinations of this parameter \cite{LQCDPi0} allow for a considerable improvement in precision, as to be described here. We are reporting an update, where we consider the most recent values of all inputs (quark masses and Z-boson mass)~\cite{Zyla}, in the perturbative region we use the running of $\alpha_s$ up to five-loop order~\cite{Baikov:2016tgj} and most importantly it is the first time that the light-quark vector current correlator at zero-momentum is known and therefore it is implemented within the method presented in this work.

\section{Hadronic contribution}
The current correlator for each flavor can be written as follows
\begin{equation}
 \Pi^{(f)}(s)=\Pi^{(f)}_{PQCD}(s)+\Pi^{(f)}_{NP}(s)+\Pi^{(f)}_{QED}(s),
\end{equation}
where on the \textit{lhs} the first contribution corresponds to the perturbative part, the second contribution is the non-perturbative one, determined using the Operator Product Expansion (OPE), and the last contribution is the lowest QED correction to the vacuum polarization. The dominant contribution to $\Pi^{(f)}(s)$ is the perturbative part.

The six different quark flavors can be organized in two sets, corresponding to the light and the heavy quarks. For the light quarks (up, down and strange), and in the massless limit, the high energy regime of $\Pi^{(f)}_{PQCD}$ is know up to order $\mathcal{O}(\alpha_s^3)$, and to order $\mathcal{O}(\alpha_s^4)$, up to a real constant. In the heavy quark sector, it is necessary to express the current correlator using both the low- and the high-energy expansion.

Turning to the heaviest quarks contribution to Eq.\eqref{eq:alpha1}, the correlator can be written in terms of the low- and the high-energy expansion. The former is given by
\begin{equation}
 \Pi_f(s)=\frac{3Q_f^2}{16\pi^2}\sum_{i=0}^\infty \overline{C}_i\Big(\frac{s}{4\overline{m}_f^2}\Big)^i,
 \label{low-correlator}
\end{equation}
where $\overline{m}_f$ is the quark mass of flavour-f in the $\overline{MS}$ scheme at a scale $\mu$, and the high energy expression is 
\begin{equation}
 \Pi(s)=Q_f^2\sum_{n=0}^\infty \Big( \frac{\alpha_s (\mu^2)}{\pi} \Big)^n \Pi^{(n)}(s).
 \label{high-correlator}
\end{equation}

The coefficients $\overline{C}_0$, $\overline{C}_1$, $\overline{C}_2$ and $\overline{C}_3$ were determined up to $\mathcal{O}(\alpha_s^3)$ in \cite{QCD1,QCD2,C2,C3}, and the terms $\Pi^{(0)}$, $\Pi^{(1)}$ and $\Pi^{(2)}$ are given in~\cite{QCD2,Pi2a,Pi2b,QCD3,QCD4,QCD5}.

In order to obtain the bottom quark contribution to  $\Delta \alpha_{\text{HAD}}(M_Z^2)$ we use Eq.~(\ref{eq:alpha1}), where $\Pi(0)$ is computed using Eq.~(\ref{low-correlator}), and $\Pi(M_Z^2)$ comes from Eq.~(\ref{high-correlator}). We obtain
\begin{eqnarray}
\Delta\alpha^{(b)}_{\text{HAD}}(M_{Z}^{2}) &=& 4\pi \alpha \left(\Pi^{(b)}(0)-\Pi^{(b)} (M_{Z}^{2}) \right) \nonumber\\
&=& (12.88 \pm 0.04) \times 10^{-4},
\label{bottomcontribution}
\end{eqnarray}
where $n_f=5$ and $\mu=10$ GeV. It is important to mention that varying $\mu$ in a range from $10$ GeV to $10M_z$ the result in Eq.~(\ref{bottomcontribution}) only changes by $0.03 \times 10^{-4}$.

For the contribution of the top-quark it is only necessary to use the low expansion to the correlator, Eq.~(\ref{low-correlator}), in Eq.~(\ref{eq:alpha1}), which gives
\begin{eqnarray}
\Delta\alpha^{(t)}_{\text{HAD}}(M_{Z}^{2}) &=& 4\pi \alpha \left(\Pi^{(t)}(0)-\Pi^{(t)} (M_{Z}^{2}) \right) \nonumber\\
&=& - (0.73 \pm 0.05) \times 10^{-4} ,
\end{eqnarray}
where $\mu=\overline{m}_t$ and $n_f=6$. We notice that the only uncertainty in this contribution is from the top-quark mass.

The next contribution is from the charm quark. Its perturbative piece follows from the Adler function approach, and it is chosen so as to minimize the uncertainty. This method takes into account a high as well as a low energy contribution
\begin{eqnarray}
 \Delta \alpha_{\text{HAD}}^{(c)}(M_Z^2)=\frac{\alpha}{3\pi}\int_{s_0}^{M_Z^2}\frac{D^{(c)}(s)}{s}ds+4\pi \alpha (\Pi^{(c)}(0)-\Pi^{(c)}(s_0)).
\end{eqnarray}

Regarding $s_0$ we choose it large enough for PQCD to be valid, but still with $s_0\ll M_Z^2$, i.e. $s_0=(9.3 \ \text{GeV})^2$. In addition $n_f=4$ and $\mu$ is taken in the range $\mu=(2-9.3)$GeV. For the low-energy expansion, we use Eq.~(\ref{low-correlator}).

The results for these contributions are discussed in detail in \cite{SB}, and are as follows
\begin{eqnarray}
\Delta\alpha^{(c)}_{\text{HAD}}(M_{Z}^{2})&=&4\pi\alpha\left(\Pi^{(c)}(0)-  \,[\Pi^{(c)}(M_{Z}^{2})]\right)\nonumber\\
&=& (79.88 \pm 0.59) \times 10^{-4}.
\end{eqnarray}

\begin{figure}[t]
	\centering
	\includegraphics[scale=0.7]{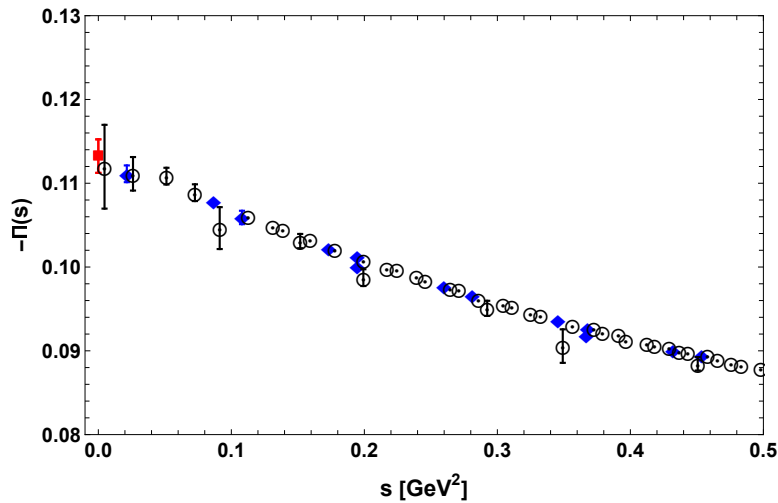}
	\caption{LQCD computation of $\Pi(s)$ \cite{LQCDPi0} with up- and down-quark contributions to the vacuum polarization function in the range $0 <s \leq 0.5$ GeV$^2$. Blue filled diamonds correspond to Fourier momenta, open black circles denote data points computed using twisted boundary conditions.  Red filled square indicates the value of $\Pi(0)$ determined from the second time momenta with $m_\pi=185$ MeV.}
	\label{fig:qed}
\end{figure}

Turning to the light-quark sector contribution, we use an entirely theoretical method involving the Adler function. The non-perturbative contribution is obtained from LQCD results, and the perturbative contribution follows from the integration on a semi-circular contour of radius $|s_0|$, avoiding the origin. This gives for $\Delta^{(uds)}_{\text {HAD}}(M_Z^2)$
\begin{eqnarray}
\Delta^{(uds)}_{\text {HAD}}(M_Z^2)&=& \Delta \alpha^{(uds)}_{\text{HAD}} (-s_0)
+ [\Delta \alpha^{(uds)}_{\text{HAD}}(s_0)- \Delta^{(uds)}_{\text{HAD}}(-s_0)]
 + [\Delta^{(uds)}_{\text{HAD}}(M_Z^2)-\Delta^{(uds)}_{\text{HAD}}(s_0)] \nonumber \\
&=& 4 \pi \alpha [\Pi^{(uds)}_{\text{LQCD}}(0)] - \Pi^{uds}_{LQCD}(-s_0)] \nonumber \\
&+& \frac{\alpha}{3 \pi}\int_{-s_0}^{s_0} \frac{D^{(uds)}_{\text{PQCD}}(s)}{s} ds + \frac{\alpha}{3 \pi}\int_{s_0}^{M_Z^2} \frac{D^{(uds)}_{\text{PQCD}}(s)}{s} ds.
\label{expressionuds}
\end{eqnarray}
In Eq.~(\ref{expressionuds}), we use $s_0=-3.5 \text{GeV}^2$ to find
\begin{equation}
\frac{\alpha}{3 \pi}\int_{-s_0}^{s_0} \frac{D^{(uds)}_{\text{PQCD}}(s)}{s} ds = (2.97 \pm 0.14) \times 10^{-4} \,,
\end{equation}
\begin{equation}
\frac{\alpha}{3 \pi}\int_{s_0}^{M_Z^2} \frac{D^{(uds)}_{\text{PQCD}}(s)}{s} ds = (125.64 \pm 0.07) \times 10^{-4}  \,,
\end{equation}
\begin{equation}
4  \pi \alpha  [ \Pi^{(uds)}_{\text{LQCD}}(0) - \Pi^{(uds)}_{\text{LQCD}}(-s_0)]= (53.49 \pm 0.40) \, \times 10^{-4} \,,
\end{equation} 
where we used the LQCD data depicted in Fig. 1, the data is from up- and down-quark contributions to the vacuum polarization function in the range $0 <s \leq 0.5$ GeV$^2$, where blue filled diamonds correspond to Fourier momenta, open black circles denote data points computed using twisted boundary conditions, and red filled square indicates the value of $\Pi(0)$ determined from the second time momenta with $m_\pi=185$ MeV. Finally, we obtain
\begin{equation}
\Delta \alpha^{(uds)}_{\text{HAD}} = (182.10 \pm 0.43) \times 10^{-4},
\end{equation}
with this value  differing substantially from the approximate value used previously  in  \cite{SB}, due to the new LQCD result for $\Pi(s)$ which is now known at the origin \cite{LQCDPi0}.\\

\section{Result}
Adding up all the contributions gives the final result
\begin{equation}
\Delta \alpha_{\text{HAD}}(M_Z^2) = (274.13 \pm 0.73) \times 10^{-4}
\end{equation}

for $n_f=6$. This result is obtained entirely from theory, as a combination of LQCD and PQCD. The main uncertainty of this approach in the past was from the value of the vector correlator at the origin. The new value of this quantity allows now for a precision result.

In order to make a fair comparison, the result of Ref.~\cite{SB}, using the same technique, is  $\Delta \alpha^{(uds)}_{\text{HAD}}=181 \times 10^-4$, with no uncertainty given, and using the LQCD information available at that time.

In the literature there is a large number of determinations of $\Delta \alpha_{\text{HAD}}(M_Z^2)$ from a variety of methods, with results in the range \cite{Table1}-\cite{Table12} 
\begin{equation}
\Delta \alpha_{\text{HAD}}(M_Z^2) = (269 - 279) \times 10^{-4},
\end{equation}

albeit with tiny individual uncertainties in each determination.

\acknowledgments

This work was supported in part by the National Research Foundation (South Africa) and by the Alexander von Humboldt Foundation (Germany).

\end{document}